# GIPPO: A Graph-based, Iterative, Printing-Path Optimization Slicer for Architected Lattices

Pierpaolo Fucile[1#], Maria Kalogeropoulou[1#], Vivek Cherian David[1], Lorenzo Moroni[1*]

Complex Tissue Regeneration, MERLN Institute for Technology-Inspired Regenerative Medicine, Maastricht University, 6229 ER Maastricht, the Netherlands.

#: Authors contributed equally to this work.

l.moroni@maastrichtuniversity.nl

## Abstract

Architected materials of significant geometric complexity offer exceptional mechanical properties that often surpass those of their constituent materials. However, their fabrication through extrusion-based 3D printing remains hindered by suboptimal printing trajectories, which is inherent to commercial slicing software. Conventional slicers produce multiple non-continuous paths that compromise fabrication time, shape fidelity, and structural integrity, particularly for thin-walled lattice structures. To address this issue, we introduce GIPPO (Graph-based, Iterative, Printing-Path Optimization), an open-source slicing platform that transforms complex lattice designs into optimized printing trajectories using graph theory. Complex lattices are converted to graph networks of nodes and edges, to derive the optimal printing trajectories through a modified version of Prim's algorithm. The resulting printing paths are translated back to Euclidean coordinates and exported as a ready-to-use G-code. We validated GIPPO's performance against conventional slicing software across six architected lattice geometries fabricated from thermoplastic polyurethane using fused deposition modeling. GIPPO-optimized constructs demonstrated superior shape fidelity with reduced local thickness deviations, no missing struts, and minimized excess material deposition compared to conventionally printed controls. Mechanical testing revealed that printing path optimization directly influences both uniaxial and out-of-plane mechanical responses, with different optimization strategies yielding distinct performance characteristics suited to specific loading conditions. Moreover, the platform accommodates both planar and non-planar printing geometries and enables fabrication of objects with varying infill patterns per layer. Our work addresses critical limitations in commercial slicing software and opens new opportunities for high-fidelity fabrication of complex architected materials.

# Introduction

Architected materials are engineered materials whose properties are primarily determined by their geometric structure and often surpass those of their constituent materials [1, 2]. The combination of tailored geometries with the fabrication freedom offered by additive manufacturing strategies, has allowed the investigation of a host of complex, new materials with outstanding mechanical properties, ranging from fatigue tolerance[3], energy absorption[4], stiffness[5] and ultra-low relative density[6], among others. For applications involving living cells, like tissue engineering[7], cultured meat[8] and engineered living materials containing bacteria or microalgae[9], architected designs offer additional control on local and global porosity, perfusion of nutrients and oxygen, and tissue growth guidance, apart from the structural integrity. Hence, platforms of rapid design conversion to printing trajectories are expected to accelerate the implementation of architected materials to various fields.

Among the different 3D printing methods that have been used to fabricate architected materials, extrusion-based approaches, such as fused-deposition modelling (FDM) and direct ink writing (DIW), enable the fabrication of architected lattices from a broad array of materials, including elastomers, ceramics[10], epoxy resins[11], multi-material composites [12] and cell-laden hydrogels[13]. This modality renders the use of extrusion printing an attractive option for the fabrication of constructs of increased geometric complexity. The main principle of extrusion printing relies on the fabrication of a given layer of a design by the deposition of an extruded filament along the design contours. To minimize fabrication time, design deviation and printing inconsistencies, the ideal printing path of a single layer should consist of one continuous line, which traces all the contours of the geometry. If the given geometry cannot be reduced to a single path, the next best solution is maximizing the length of the sub-trajectories and, by extension minimizing the number of small paths.

However, translation of architected lattices to feasible, if not optimal, printing trajectories is not a straightforward process and greatly depends on commercial slicing software. Most commercial systems typically include a library of certain basic infill geometries (*e.g.*, woodpile, triangles), which is limited and software dependent [14]. Even when more complex geometry designs are supported, the slicing software typically produces multiple non-continuous printing paths that compromise the fabrication time and shape fidelity, as well as the structural integrity of the final piece. Additionally, custom lattice designs can only be fabricated by commercial systems after tedious manipulation of the fabrication parameters and may still result in printing trajectories with multiple inconsistencies.

While some research has been done on the printing path optimization for extrusion printing, most approaches focus on the fabrication of solid three-dimensional bodies [15]. Even when lattice constructs are considered, their struts typically consist of multiple walls, exceeding the thickness of the extruded filament. Recently, promising attempts to introduce path planning and toolpath corrections in embedded printing and melt-electrowriting, respectively, have been presented, further illustrating the increasing need for similar platforms for the reliable fabrication of architected materials [11, 16]. Furthermore, in the construction field, it has been used to generate concrete components and minimize printing time by optimizing the order of printing of each component [17].

Here, we introduce a Graph-based, Iterative, Printing-Path Optimization slicing software (GIPPO) for the fabrication of thin-walled architected lattices. Complex lattices are first designed using parametric visual programming, segmented into their constituent nodes and edges, and subsequently transformed to undirected graphs. The optimized printing paths are extracted using a modified version of Prim's Algorithm, which traditionally derives the minimum spanning tree from a connected graph network following a greedy approach [18, 19]. The main difference is that GIPPO does not allow for branching paths, and unlike typical applications (e.g., travelling salesman problem, swarm intelligence algorithms, Dijkstra's algorithm), our main requirement is that every edge must be visited, and hence every node must be completed. The solution is then exported as a sequence of nodes grouped in distinct printing paths, which are converted back to the Euclidean coordinates matching the nodes of the original geometry. Despite its complexity, the graph-based algorithm is meant to be used by a wide range of researchers through user-friendly interfaces, aiming to boost architected matter fabrication. To demonstrate the applicability of GIPPO, the optimized printing paths are translated to a G-code script and used for fabricating architected lattices from thermoplastic polyurethane using FDM printing. We investigated the impact of the printing path generation on the shape fidelity of the lattices compared to controls fabricated using conventional slicing software. The effect of the printing path generation on both the uniaxial and out-of-plane mechanical responses of the lattices was also assessed. Finally, we showcased how GIPPO can be used for the printing optimization of non-planar lattices and for the extreme case of three-dimensional objects with different infill geometry per layer.

## Materials and Methods

**Comprehensive platform for complex scaffolds generation**

Our free-to-use software was developed in Rhinoceros® Grasshopper (*Version 8.0, Robert McNeel & Associates, Seattle, Washington*), and MATLAB (*Version R2024b, The MathWorks Inc., Natick, Massachusetts*). It comprises three main blocks, namely: i) design generation and conversion to graph; ii) graph-theory-based trajectory optimization; and iii) post-processing and *firmware-agnostic* 3D printing file generation. In particular, Grasshopper takes care of block *i* and *iii*, while block *ii* is handled in MATLAB environment.

*Design generation*

The generation of the designs was done through a custom-made Grasshopper script which was developed to process parametrically multiple geometrical features with a high degree of control. The input parameters include: (i) starting geometry, which can be generated either through standard tools or imported from an external STL (i.e., Standard Tessellation Language) file; (ii) layer thickness (i.e. the distance center-to-center between two consecutive layers); (iii) angle of rotation between layers; (iv) strand distance (i.e., the distance center-to-center between two adjacent lines within the same layer); (v) porosity, through a library of *ready-to-use* porosity types (i.e., infills); (vi) number of layers; (vii) printing surface, which in case of 3D printing on a non-planar surface can be imported for the software to project the sliced structures onto it; and (viii) the building angle, which in case of a non-planar printing surface is the normal vector to the surface.

Once all the design parameters are set, the starting geometry is virtually *sliced* according to the layer thickness. The outer contours of each of the layers are extracted, "filled" with the selected infill geometry and rotated if required. In particular, for each layer, at least double its surface is filled with porosity and then cropped with the contours.
For this specific study, six geometries were selected from the built-in atlas, namely Honeycomb grid, Voronoi grid, Snub Square and Penrose tiles, and two auxetic geometries, namely Arrowhead and Re-entrant Honeycomb. All geometries were designed as polylines. In particular, the tiling geometries were designed through a Grasshopper plugin, namely Parakeet.
A pipe thickness similar to the filament thickness of the printing system was applied to the designs, to evaluate the expected pore area and dimensions, as shown in **Figure 1**. The pattern-specific parameters used in the design are specified **Table 1**. All parameters were set to achieve a final value of pores area within the range of 50-70% of the layer surface (i.e., 10x10 mm).

Table 1. List of distinctive topological parameters and pores density for each fabricated geometry.

| Geometry | Pattern-specific parameter | Pores area [%] |
|---|---|---|
| Honeycomb | Hexagon radius | 75 |
| Voronoi | Points density per mm$^3$ | 69 |
| Penrose tiling | Grid cells size | 59 |
| Snub Square tiling | Grid cells size | 55 |
| Arrowhead | Vertical and horizontal lengths of the unit cell [20] | 55 |
| Re-entrant Honeycomb | Vertical and horizontal lengths of the unit cell [20] | 63 |

The 10x10 mm grids were designed as 1-layer structures for microscopy analysis. The same lattices were generated for G-code universality assessment to be fabricated with a different 3D printing system. 2-layers grids were considered for tensile tests. In particular, 10x3 mm handles were included in the Grasshopper design. Furthermore, 2-layers cylindrical structures with a diameter of 48 mm were designed, optimized, and fabricated for ball burst tests. In this case, additional contours were included as handles for better adhesion with the testing hardware.

To prove the full functionalities of our software towards real experimental applications (i.e., 3D complex scaffolds for regenerative medicine), a concave cylinder with a base diameter of 10 mm and height of 4 mm was sliced, selecting Honeycomb as infill with a rotation angle between layers of 45 degrees. Given the nature of the starting volume, each one of the 28 layers was different from one another. Non-planar features were validated through the generation of an optimized 12x4.2 mm cylindrical scaffold built onto a virtual spherical surface. All the structures were sliced with a layer thickness of 0.148 mm.

Each scaffold design is then processed through a python code within Grasshopper in which a node label is assigned to each vertex of each layer of the structure (**Figure 1**). The outcome of this process (i.e., the pairs of nodes associated to the segments of the geometry) is finally exported towards the optimization step.

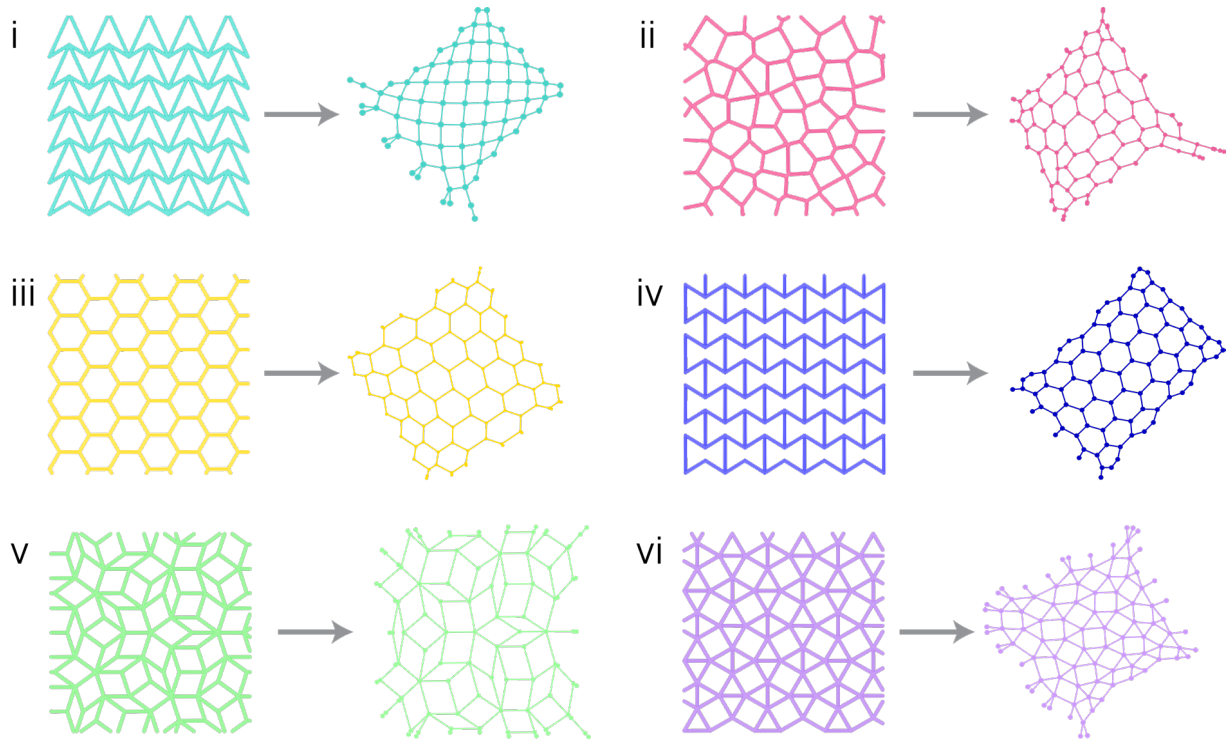

**Figure 1: The six geometries selected for this study.** Arrowhead **(i)**, Voronoi **(ii)**, Honeycomb **(iii)**, Re-entrant Honeycomb (RE Honeycomb) **(iv)**, Penrose **(v)**, Snub Square **(vi)**. At the left side of the arrows the 3D models generated with GIPPO's atlas are reported. At the right side the representative graph generated in MATLAB.

*Graph-theory-based trajectory optimization*

The pseudocode of our algorithm is provided in Algorithm 1.

**Algorithm 1: The pseudocode of the GIPPO algorithm, developed in MATLAB.**

---

**Input:** Nodes pairs from Grasshopper $n_i$ $n_j$
**Output:** Optimized arrangement of nodes to form a set of "longest possible" segments

1. Generate graph **G**;
2. Set all nodes $n_i$ to *incomplete*;
3. **for** k=1:number of iterations
4.     **while** all nodes are *incomplete*
5.         **if** new segment
6.             select a random node $n_i$;
7.         **else** select current growing segment's extreme vertices;
8.         check neighbour nodes;
9.         **if all** neighbour nodes ∈ previous segments
10.             $n_i$ ∈ completed nodes;
11.             segment complete;
12.             **return**
13.         **else** include neighbour with highest (or smallest) distance from $n_i$;
14.     scoring solutions $s_k$= set of segments;
15. select highest score solution $s_{best}$;
16. sort segments of $s_{best}$ from longest to shortest;
17. **return** $s_{best}$;

---

The nodes pair are imported in MATLAB to generate a graph $G$. The lengths of the segments in mm are imported as *weights* of the graph edges and used by the algorithm to determine the direction of growth of the paths.

The algorithm is iterative and produces 500 solutions, one per iteration. Nonetheless, MATLAB's Parallel Computing Toolbox is employed to decrease computational times by 75-80 %.

Each solution is a set of paths, namely an arrangement of the design segments, based on a furthest (or closest) neighbor node approach, until all nodes are *completed*. A node is considered completed when all the edges associated to it are part of a path.

Each path is generated starting from a random *incomplete* node. For every new node in the path, the neighbors of both extreme vertices are analyzed. GIPPO allows for a *maximum* or a *minimum* based optimization. In the first case, the neighbor with the longest distance from the current node is included in the path, and vice versa for the minimum optimization. When a path cannot grow anymore (i.e., when all the neighbors of the extreme vertices are completed nodes) the growth stops and the algorithm starts a new path.

After all the solutions are generated, each one of them is scored according to the following scoring function (Equation 1):

$$score = \sum_i \frac{length_i * NumEdges_i}{TotalLength * TotalPaths} \qquad 1$$

Where $length_i$ is the length (in mm) of the $i^{th}$ path of the solution, $NumEdges_i$ is the number of segments of the $i^{th}$ path, while $TotalLength$ and $TotalPaths$ are the total length in mm and the total number of paths of the solution, respectively. The highest score solution is then selected and its paths sorted from longest to shortest. The solution is finally exported for G-code generation.

The following indicators were analyzed to evaluate GIPPO's robustness and the quality of the generated solutions:

- Maximum scores over 500 algorithm executions;

- Long-to-short (LTS) paths ratio over 500 algorithm executions, as defined per the following equation (Equation 2):

$$LTS\% = \frac{\sum_i LengthLong_i}{TotalLength}\% \qquad 2$$

Where $LengthLong_i$ is the length (in mm) of the $i^{th}$ long path of the solution. A path is considered long if its number of nodes $n$ is greater than or equal to 5.

- Optimization efficiency (OE), which was evaluated to assess the increase in paths quality between the best (*i.e.*, best score) and worst (*i.e.*, lowest score) solutions, for both maximum and minimum approaches. OE was defined as per the following equation (Equation 3):

$$OE\% = \sum_i \frac{LengthLong_i * NumEdgesLong_i}{TotalLength * TotalEdges}\% \qquad 3$$

Where $NumEdgesLong_i$ is the number of edges (i.e., segments) of the i-th long path of the solution, and $TotalEdges$ is the total number of edges of the solution.

*Post-processing and 3D printing file generation*

Optimized trajectories are re-imported in Grasshopper and associated to the starting nodes of the geometry. In case of tensile tests, customized handles can be included for each layer, or contours can be added to the printing structure if needed.
Our G-code generation platform currently allows for the generation of Marlin compatible codes and *.nc* codes, which run on specific pneumatic-based 3D printing systems.
GIPPO has a high degree of control over many printing parameters, e.g. flow rate, pre-flow, post-flow, retraction, etc. In particular, extrusion values (*i.e*, "E-value", the amount of material to be extruded for each segment) are derived from the following equation (Equation 4), which is an adaptation of typical E-values equations used in the 3D printing community:

$$Evalue_i = k \frac{4 * length_i * LT * EM * d_n}{\pi * f_d^2} \qquad 4$$

Where $LT$ is the layer thickness, $EM$ the extrusion multiplier, $d_n$ the nozzle diameter, $f_d$ the filament diameter (i.e., typically 1.75 mm), and $k$ a correction factor that accounts for the non-cylindrical shape (i.e., the theoretical one) of extruded filaments, which are rather elliptical. Additionally, the software automatically recognizes short paths (i.e., less than 4 segments), and can tune the printing speed to allow for a better extrusion and adhesion between adjacent filaments. In case of pneumatic-based systems, pressure and flow values are assigned to each segment.

**Lattice fabrication**

In order to validate our software in terms of quality of the optimization process and ease of use for a wide range of researchers with different backgrounds, we selected an entry level low-cost Fused Deposition Modelling (FDM) 3D printer (Creality Ender 3 S1 Pro, Creality, Shenzhen, China) equipped with a 0.20 mm needle (FabConstruct Nozzle MK8 0.20 mm, Fabistron GmbH, Germany). A commercial Thermoplastic Polyurethane (TPU, RealFlex, Real Filament, The Netherlands) was used to 3D print the samples. A free-to-use slicing software (Ultimaker Cura, Version 5.6, Ultimaker, The Netherlands), hereinafter referred as "control slicer", was used to generate non-optimized G-codes. Commercial slicers do not typically include complex pore design. For this reason, our lattices were exported as .stl files to be processed by the control slicer. To allow for proper material deposition (*i.e.,* following the design structure), the geometry had to be defined as a "wall structure", with a 0% infill.

For each sample type (i.e., microscopy analysis, tensile and ball burst tests, and concave models), printing parameters were kept the same for the optimized approach and the controls, and are presented in Table S1. Differences in printing trajectories and movements were analyzed by recording the printer's movements. A glass printing bed was placed on top of the built-in one, up to a height sufficient for a camera (ArduCam 120fps Global Shutter USB Camera Board 1MP OV9281 UVC, M12 Lens, Arducam, Nanjing, China) to be on focus on the needle tip while printing. Videos were processed and analyzed with Adobe Premiere (Adobe, USA).

In order to assess the correct generation of *firmware-agnostic* printing codes deriving from the same optimized design, equivalent parameters were controlled and set on the pneumatic printing code generation platform of our software, and 10x10 single layer lattices were printed with a pneumatic-based system (Bioscaffolder, SysENG, Germany). In particular, maximum- and minimum-based lattices were printed using Polycaprolactone (PCL, $M_w$=80kDa, Sigma Aldrich, Germany) and a 27G needle (DL technology, Haverhill, MA, USA). Equivalent printing parameters are reported in Table S2.

**Microscopy and Printing Fidelity Analysis**

All the single-layer lattices (i.e., TPU and PCL structures) were imaged with a stereomicroscope (SMZ25, Nikon instruments, USA) with a darkfield ring illuminator (Nikon Instruments, USA) to evaluate fiber thicknesses and printing fidelity. In particular, the PCL geometries were imaged to qualitatively assess the validity of the universal G-codes. All images were processed in ImageJ.

The assessment of the printing fidelity of the three different printing approaches was conducted using a local thickness mapping method. More specifically, stereomicroscopy images of the printed scaffolds (*n=3*) were imported to ImageJ and transformed into 8-bit images. Subsequently, the threshold of each image was adjusted to create a mask including only the scaffold and a map of the local thickness of each structure was made using the Local Thickness analysis. The histogram of each map was exported and converted to width values. The local thickness histograms were plotted using Origin 2018.

The mean and standard deviation of each histogram were calculated using the following formulas:

$$\mu = \frac{\sum(x_i * f_i)}{\sum f_i}, \qquad i = 1, \ldots, 256 \qquad\qquad 5$$

$$\sigma = \sqrt{\frac{\sum f_i * (x_i - \mu)^2}{\sum(x_i * f_i)}}, \qquad i = 1, \ldots, 256 \qquad\qquad 6$$

where $\mu$ is the mean, $\sigma$ is the standard deviation, $x_i$ is the $i^{th}$ unique, thickness value and $f_i$ is the corresponding frequency at which this value occurs during the local thickness mapping.

Finally, the combined mean and standard deviation of all replicates per condition were calculated as follows:

$$\mu_{combined} = \frac{\sum(n_j * \mu_j)}{\sum n_j}, \qquad j = 1, 2, 3 \qquad\qquad 7$$

$$\sigma_{combined} = \sqrt{\frac{n_j * \sigma_j^2 + n_j(\mu_{combined} - \mu_j)^2}{\sum n_j}}, \qquad j = 1, 2, 3 \qquad\qquad 8$$

where $n_j$ is the number of thickness measurements in replicate $j$, given by $n_j = \sum f_i$.

**Trajectories analysis**

Additional differences in terms of printing fidelity were investigated from a trajectory point of view. To do so, a custom-made MATLAB code was developed to extract printing paths from control G-codes. These trajectories were compared to the optimized ones in terms of total printing distances (i.e., a measure of extruded material) to the nominal ones, and "GIPPO score", as per a modified version of Equation 1 which includes a correction factor that keeps into account the increase of final segments to the nominal ones (i.e., eventual print overlaps within the same layer).

**Tensile Tests and Analysis**

The mechanical performance of the optimized and non-optimized lattices was assessed in uniaxial tension using a TA Force equipped with a load cell of 444 N. The samples were pulled to an ultimate strain of 300% with a strain rate of 0.1 mm/s. If the samples fractured before the maximum strain limit, the measurement was stopped. Five samples were tested per condition (n=5). Data was collected with a rate of 100 datapoints per second.

Stress ($\sigma$) and strain ($\varepsilon$) data were calculated from the force ($F$) and displacement ($\Delta l$) measurements using the following formulas:

$$\sigma = \frac{F}{A} \qquad\qquad 9$$

$$\varepsilon = \frac{\Delta l}{l} \qquad\qquad 10$$

where $A$ the area of the sample cross-section and $l$ the sample length.

The stress-strain curves were fitted to a second-order power series with the general formula of:

$$\sigma = a \cdot \varepsilon b + c \qquad\qquad 11$$

where $a, b, c$ the model coefficients. The fitting was performed using MATLAB 2022b. The goodness of fit was evaluated by ensuring that $R^2 > 0.99$. Specific exclusion rules were applied on each curve, to define the upper strain limit for the fitting, to avoid including post-fracture stress values that would cause a deviation of the fitted curve. The elastic modulus was calculated as the inclination of the linear region of the stress-strain curve in the 5%-7% strain range. The fracture points were calculated as the first stress value with a deviation equal to or larger than 10% from the fitted stress, for the same strain, using a custom-made MATLAB script. The toughness to fracture was calculated as the area under the fitted curve of the stress-strain data, until the fracture strain, using integration. Results were plotted using GraphPad 10 as average values ± standard deviation.

**Ball Burst Tests**

The out-of-plane performance of the printed constructs was evaluated using a ball-burst test. A custom accessory for the TA mechanical test frame was designed and fabricated from stainless steel (xometry, China) following the ASTM-D3787-16-2020**.** All tests were conducted using a load cell of 450 N at a …. strain rate. The tests were terminated at a maximum displacement of 30 mm from the bottom surface of the samples, or at a lower strain in case the samples ruptured. Three samples were tested per condition (*n=3*). The local deformation modes were assessed using a camera (ArduCam 120fps Global Shutter USB Camera Board 1MP OV9281 UVC, M12 Lens, Arducam, Nanjing, China) fixed under the sample stage.

The shape formability of the samples was assessed by the load-displacement curves. The bursting strength was determined by the load level at which the samples exhibited their first tear. The bursting displacement was defined as the corresponding displacement level of the first tear.

**Statistical Analysis**

A one-way or two-way ANOVA with a post-hoc Tukey's multiple test was used to identify any statistically significant differences across the tested conditions. Statistical analysis of the data was performed in GraphPad10. Statistical significance: *: P<0.05, **: P<0.01, ***: P<0.001, ****:P<0.0001.

# Results

## Algorithm robustness and paths construction

The difference between the best and worst score solution of a GIPPO execution is reported in **Figure 2 (i-iii)**, and **Figure S1**, where an overview of all the cases is reported for each condition, including the one in **Figure 2 (ii-iii)**. For each geometry and each optimization method (i.e., maximum or minimum based), the generated paths were reconstructed and plotted on top of the original lattices. In particular, from each score analysis, the paths associated to the highest and lowest score (i.e., the red and green dots in **Figure 2 (i)**, respectively), were reconstructed and their paths divided into "long" (i.e., more than 15 nodes), "medium" (i.e., between 5 and 15 nodes), and "short" (i.e., less than 5 nodes). GIPPO succeeds in rearranging the starting grid segments into the longest paths possible, thus reducing short segments and hence possible weak points in the final structure, with tangible differences between best and worst solutions, as confirmed by the Optimization Efficiency (OE) analysis (**Figure 2 (iv)**). This parameter kept into account each path's length in mm and "graph edges" and was normalized by the total of both. As expected, "best" approaches always performed better, for both minimum and maximum optimizations. Furthermore, maximum optimization proved to be more effective for all the geometries but the arrowhead, in which the minimum-based approach provides a slightly higher efficiency value.

Given the random nature of GIPPO (i.e., each path of each iteration is built upon a random seed), the robustness of the algorithm over 500 independent executions was assessed to confirm the quality of the generated solutions. In particular, the maximum score for each geometry and each optimization method was evaluated (**Figure 2 (v)**), as well as the ratio of long over short (LTS) paths of the solutions (**Figure 2 (vi)**), which is an indicator of the coverage of long paths on the entire lattice. Scores were robust over the iterations, as shown by the standard deviation which was in the order of 1-6% on all geometries and all conditions, except for the arrowhead, in which the minimum optimization presented a nonvariant average. A similar result was evidenced in the LTS analysis, where the arrowhead minimum optimization presented an almost-zero standard deviation. This showed how for specific cases where the lattice is "simpler" (i.e., with a more uniform lengths distribution), a minimum based optimization can be more effective, as evidenced by the OE analysis as well. All the other geometries and conditions varied between 1% and 7%, thus proving the robustness of GIPPO as an optimization software for complex geometries.

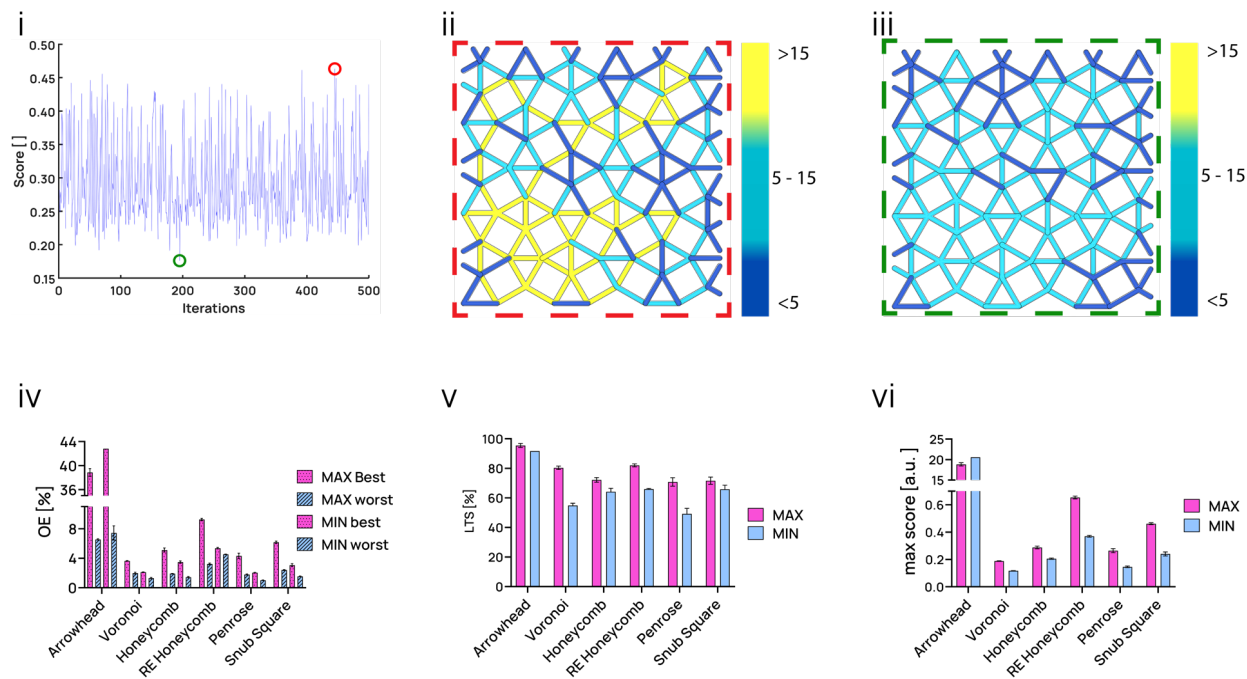

**Figure 2. Robustness and solutions quality analyses on the GIPPO algorithm. (i)** The score plot comprising the score of each of the 500 iterations. **(ii)** An example of best (i.e., highest score of the algorithm execution) vs **(iii)** worst (i.e., the lowest score) solutions. The generated paths were reconstructed for graphical assessment. The colour bars represent the number of nodes for each category (i.e., long, medium, and short paths). **(iv)** Optimization Efficiency (OE) analysis to assess the effectiveness of the optimization process (n=3). **(v)** Long-to-Short (LTS) paths evaluation over 500 independent executions of the GIPPO algorithm. **(vi)** Analysis of the best score over 500 independent executions of the GIPPO algorithm. Figure abbreviations: Max: Maximum-based Optimization, Min: Minimum-based Optimization.

## Printing Fidelity

The printing fidelity of the optimized and control trajectories was compared to their original CAD designs, in order to identify fabrication-specific deviations of each printing strategy (**Figure 3**, **Figures S1-S5**). The thickness mapping revealed substantial differences in the fidelity of the fabricated lattices. More specifically, for the arrowhead design, points of relatively increased local thickness were observed in both the minimum and maximum-based optimization results (**Figure 3 (vi-vii)**). Comparison of the arrowhead histograms against theoretical values showed that the maximum-based optimization yielded the optimal results with an average thickness of 403.9 µm, compared to the theoretical value of 245.08 µm (**Figure 3 (ix)**). A moderately higher, albeit statistically significant, thickness was measured for the minimum optimized samples (470.68 µm) (**Figure 3 (x)**). Conversely, the control samples consistently exhibited regions of excessive local thickening across all designs, as indicated by the highlighted white spots in the analysis maps (**Figure 3 (viii)**) and their dramatically increased average thickness of 928.62 µm (**Figure 3 (x))**.

The analysis of the rest of the geometries revealed that the maximum-based optimization yielded lattices with an average thickness that was consistently closer to the theoretical average thickness of the CAD structures (**Figure 3 (x), Figures S2-S5**). The minimum-based optimization outperformed its maximum counterpart only in the case of the snub-square geometry, however the average thickness difference between the two conditions was approximately 24 µm (max: 448 µm, min: 424.5 µm). Surprisingly, the control conditions across all geometries showed higher average strut thickness values and exhibited obvious printing inconsistencies.

Overall, the printing fidelity assessment revealed a prevalence of the maximum-based optimization method, by the acquisition of lattices with the minimum deviation from the original designs. Moreover, the minimum-based optimization approach resulted in constructs with significantly lower thickness values compared to the control conditions. These findings highlight the importance of introducing a G-code optimization step before proceeding with the fabrication of complex geometries.

The analysis of printing movements supported these results, as shown in **Video S1**, where a representative example (*i.e.,* Snub Square maximum optimization vs control slicer side by side) is reported. The local thickening areas highlighted for control samples during the printing fidelity assessment were evident and can be localized in areas where the extruder would pass on already printed segments. Furthermore, this analysis highlighted a higher presence of oozing and small strands over non-printing trajectories in the case of GIPPO-optimized structures.

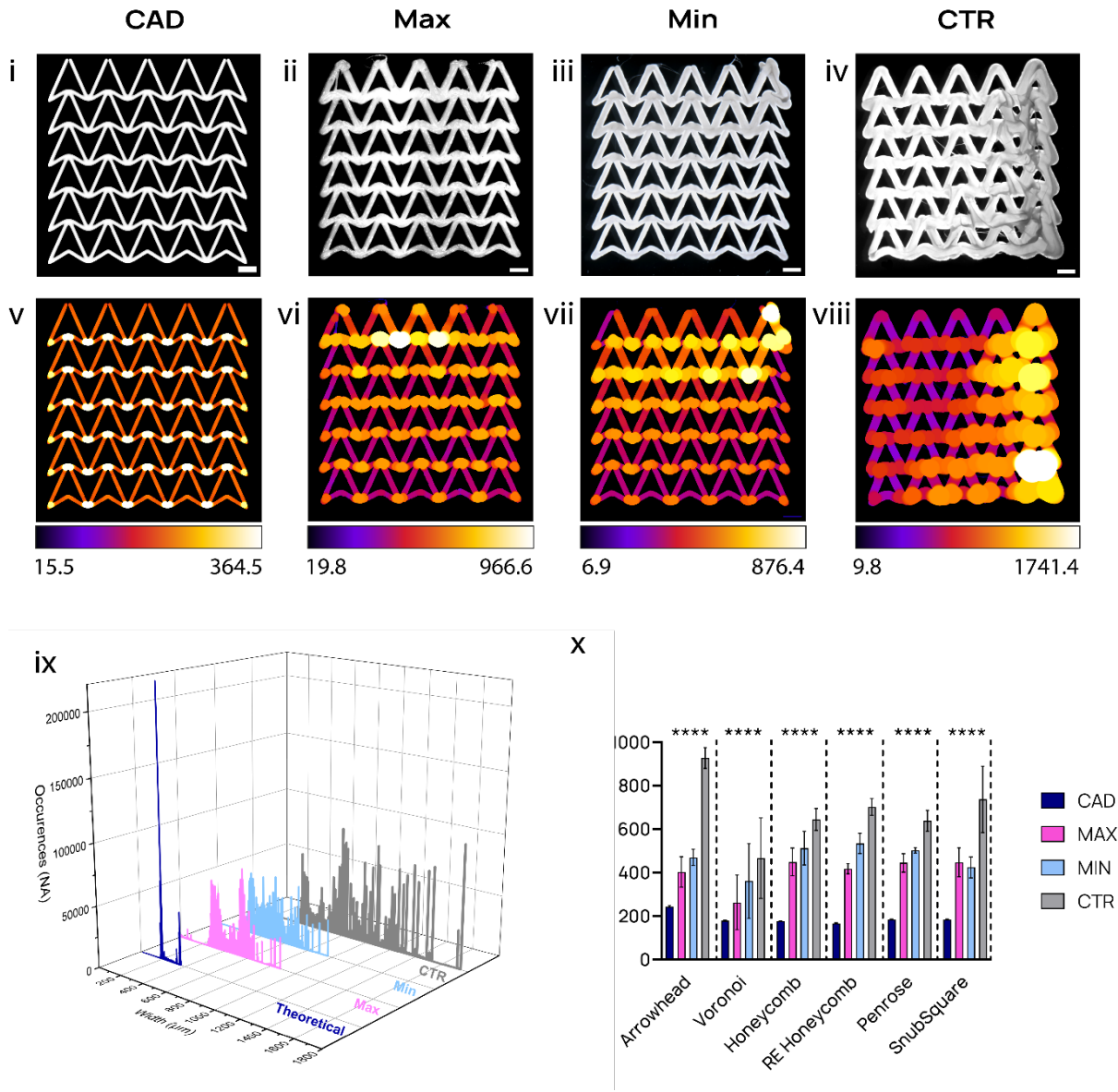

**Figure 3. Printing fidelity assessment of the two optimization approaches (min and max), compared to the theoretical design and the control slicing software for the Arrowhead geometry. (i)** CAD model and **(ii-iv)** stereomicroscopy images and their corresponding thickness maps **(v-viii)**. Scale bars: 500 µm. Minimum and maximum values are given in µm. Color bars indicate the thickness variance within each sample in µm. **(ix)** Thickness distributions calculated from the thickness maps for each condition. **(x)** Strut thickness of the control and optimized constructs using the minimum and maximum approach compared to the theoretical strut thickness of the CAD designs. The four stars (****) on top of each design group indicate that all comparisons within this design were significantly different with a significance of P<0.0001. Figure abbreviations: CAD: computer-aided design, Max: Maximum-based Optimization, Min: Minimum-based Optimization, CTR: Control.

## Trajectory reconstruction and analysis

Since the printing fidelity analysis highlighted thickening regions in the control samples, additional analyses on the way the trajectory is handled and processed right before fabrication were performed, to check for eventual double segments, i.e. segments in which the printer extrudes on already printed paths.

The custom-made MATLAB script allowed us to reverse engineer each geometry starting from the G-code, as presented in **Figure 4 (i-iii)** and **Figure S6**. Due to the internal handling of stl files in the control slicer, the final trajectories differed from the nominal ones, thus resulting in overlapping paths and detached designed portions. This explains the thickenings and the higher incidence of disconnected paths, hence eventual holes, in the control lattices, as evidenced in the control samples of **Figure 3** and **Figure S2-S5 (ii-iv)**.

To further validate the presence of overlapping paths, GIPPO scores and total trajectory length were analyzed (n=3). In particular, Equation 1 was modified to include a correction factor, namely $EdgesTotReal/EdgesTotNominal$, which aims to penalize trajectories with a total number of edges higher than the nominal one (i.e., overlapping paths). GIPPO score was evaluated to assess the quality of the control slicer solution, compared to the optimized one (**Figure 4 (iv)**). Optimized trajectories performed better than non-optimized ones for all geometries except for the Voronoi and the Penrose. This depends on (i) GIPPO score rewarding very long segments, hence in case of very long overlapping segments they might still result in higher scores; and (ii) control slicers subdividing edges in additional points. It was also noted that for the same set of parameters, the control slicer generated always the same G-code, thus resulting in no deviation between samples of the same condition. To further support these results, total paths length in millimeters (i.e., the total travel distance) was evaluated (**Figure 4 (v)**). Also in this case, control paths presented a 2-fold increase in the extruding movements (i.e., the "G1" commands), meaning a higher chance of overlapping paths. The only exception was the Honeycomb, with a 1 mm increase of total length compared to the nominal one. Though, this solution was considered poor through the previous GIPPO score assessment.

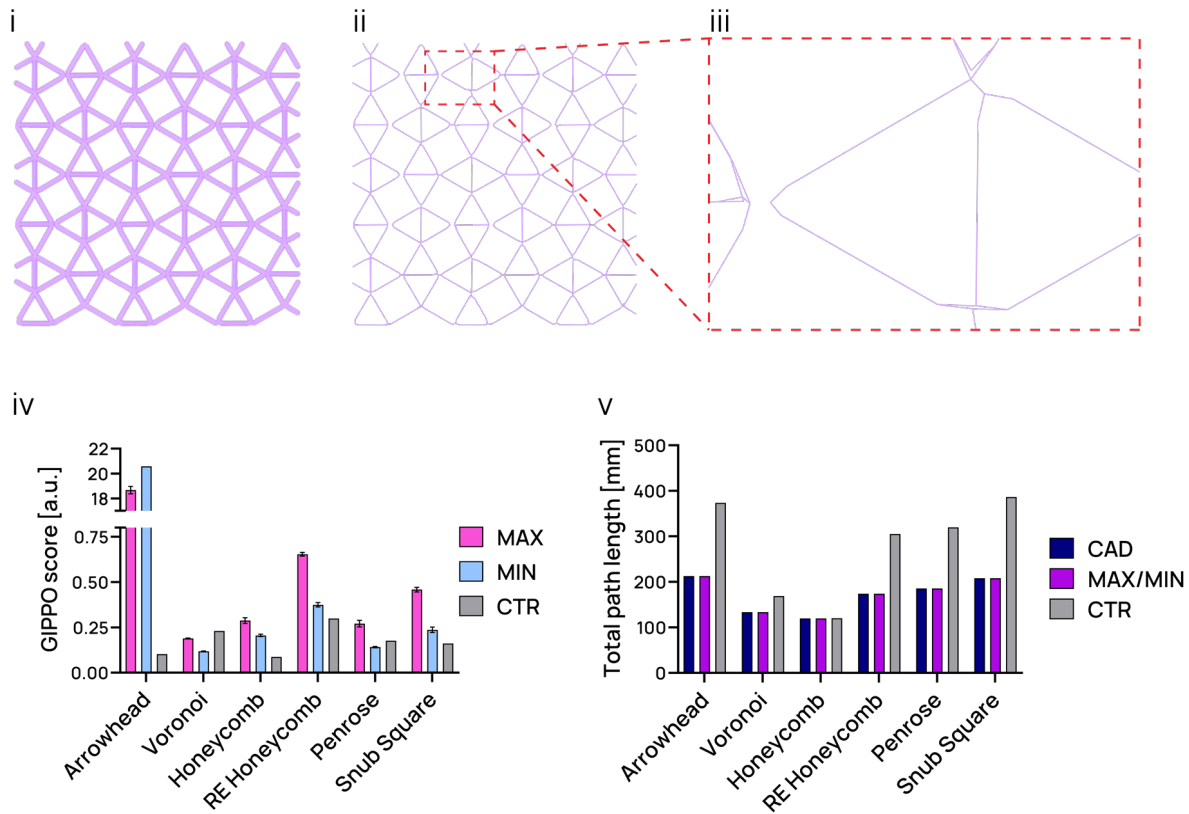

**Figure 4: Reverse engineering and analysis of the control slicer G-codes. (i)** Reconstructed 3D grid of a Snub Square lattice starting from the control G-code. **(ii)** Analysis of the polylines associated to the 3D grid showing the reprocessing performed by the control slicer. **(iii)** Focus on structural defects deriving from the reprocessing. **(iv)** Differences on GIPPO score and **(v)** Total path length (i.e., the distance travelled by the extruder while printing) between optimized and non optimized trajectories. Total path length is compared to the nominal one, deriving from the CAD model. Figure abbreviations: CAD: computer-aided design, Max: Maximum-based Optimization, Min: Minimum-based Optimization, CTR: Control.

## Mechanical Response Under Uniaxial Tensile Loading

The performance of the optimized and control samples was further assessed under uniaxial loading. Overall, the generation of the printing path was found to have a direct effect on the mechanical performance of the lattices (**Figure 5**). The conventionally generated printing paths yielded stiffer constructs compared to their optimized counterparts. This phenomenon was more pronounced for the arrowhead, snub square and Penrose geometries, as can be observed by the steeper elastic regimes of their stress-strain curves (**Figure 5 (i), (v-vi)**). Interestingly, the honeycomb, re-entrant honeycomb and Voronoi lattices exhibited a similar stress-strain curve, at least for up to 10% strain, for all printing path generation conditions (**Figure 5 (ii-iv)**).

These differences were reflected in the tensile moduli of the lattices, with the arrowhead and snub square designs showing the most striking differences. Using the minimum-based optimization approach, the modulus of the arrowhead was reduced by ~34%, compared to the conventionally printed samples. Similarly, both minimum- and maximum-based optimization approaches yielded elastic moduli reduced by ~37% and ~43%, respectively, compared to their control counterparts. Additionally, a lower tensile modulus was also obtained by optimizing the printing of the Penrose lattices (~34% and ~30%, for the minimum and maximum optimization, respectively). Finally, similar tensile moduli values were measured for the rest of the designs.

Moreover, the printing path generation strategy was found to considerably affect the fracture strain of the different geometries (**Figure 5 (viii)**). It was, overall, observed that the optimized constructs exhibited higher fracture strain values compared to the conventionally generated ones. This trend was more pronounced for the re-entrant honeycomb, arrowhead, Penrose and snub square geometries.

More specifically, the maximum-based optimization resulted in a 3-fold increase of the fracture strain in the re-entrant honeycomb lattices (154.75±30.24%), compared to the control condition (52.09±22.94%). A similar increase was revealed also for the arrowhead design, with both optimization methods showing a more than 3 times higher fracture strain compared to the controls (max: 68.74±3.85%, min: 77.50±14.92%, control: 21.14±5.85%). Interestingly, for the snub square lattices, the maximum-based optimization yielded structures with a similar fracture strain to the control (28.07±11.57% and 24.91±5.00%, respectively), while the minimum optimization significantly enhanced the fracture strain of this geometry to 57.02±9.17%. The opposite trend was observed for the Penrose geometries, with the minimum optimization and control conditions both fracturing at an average strain of 20.06±0.19% and 19.97±0.002%, while their maximum-based optimized counterparts fractured at 65.08±8.57% strain. Finally, the Voronoi and honeycomb geometries exhibited similar fracture strains regardless of the printing path generation method.

The fracture stress calculation further highlighted the effect of the printing trajectory on the mechanical performance of different geometries (**Figure 5 (ix)**). In most geometries, the conventional printing path generation resulted in lattices with considerably lower fracture stress values compared to the optimized conditions. The maximum-based optimization yielded higher fracture stress values in the arrowhead, honeycomb, re-entrant honeycomb and Penrose geometries (2.21±0.26 MPa, 1.87±0.19 MPa, 1.54±0.18 MPa and 1.45±0.12 MPa, respectively), while the minimum-based optimization resulted in a higher fracture stress value for the Voronoi and snub square designs (1.93±0.30 MPa and 1.47±0.18 MPa, respectively).

Finally, the toughness of each condition was measured to investigate the energy that could be absorbed before the onset of mechanical failure (**Figure 5 (x)**). The control conditions demonstrated a considerably lower toughness regardless of the lattice design, further illustrating the importance of optimizing the printing path during the fabrication of architected lattices. A striking toughness increase was observed for the re-entrant honeycomb lattices when they were fabricated using the maximum-based optimization method (174.69±45.57 MPa) compared to the control condition (33.27±21.77 MPa). A 5-fold increase was also observed in the toughness of both the maximum- and minimum-based optimized arrowhead constructs (113.94±18.05 MPa and 101.62±22.93 MPa, respectively). Similarly to the fracture stress results, the maximum optimization prevailed for the honeycomb and Penrose geometries, while the minimum for the snub square case. Furthermore, the minimum-based optimized Voronoi lattices outperformed both the maximum and the control condition.

Collectively, the tensile test results illustrate the direct effect of the printing path generation method on the mechanical performance of architected lattices. Moreover, it was observed that, depending on the geometry, one of the two proposed optimization methods yielded the best results. Hence, it is important to consider both approaches, as the performance of the final structure can be undermined by the use of sub-optimal printing trajectories.

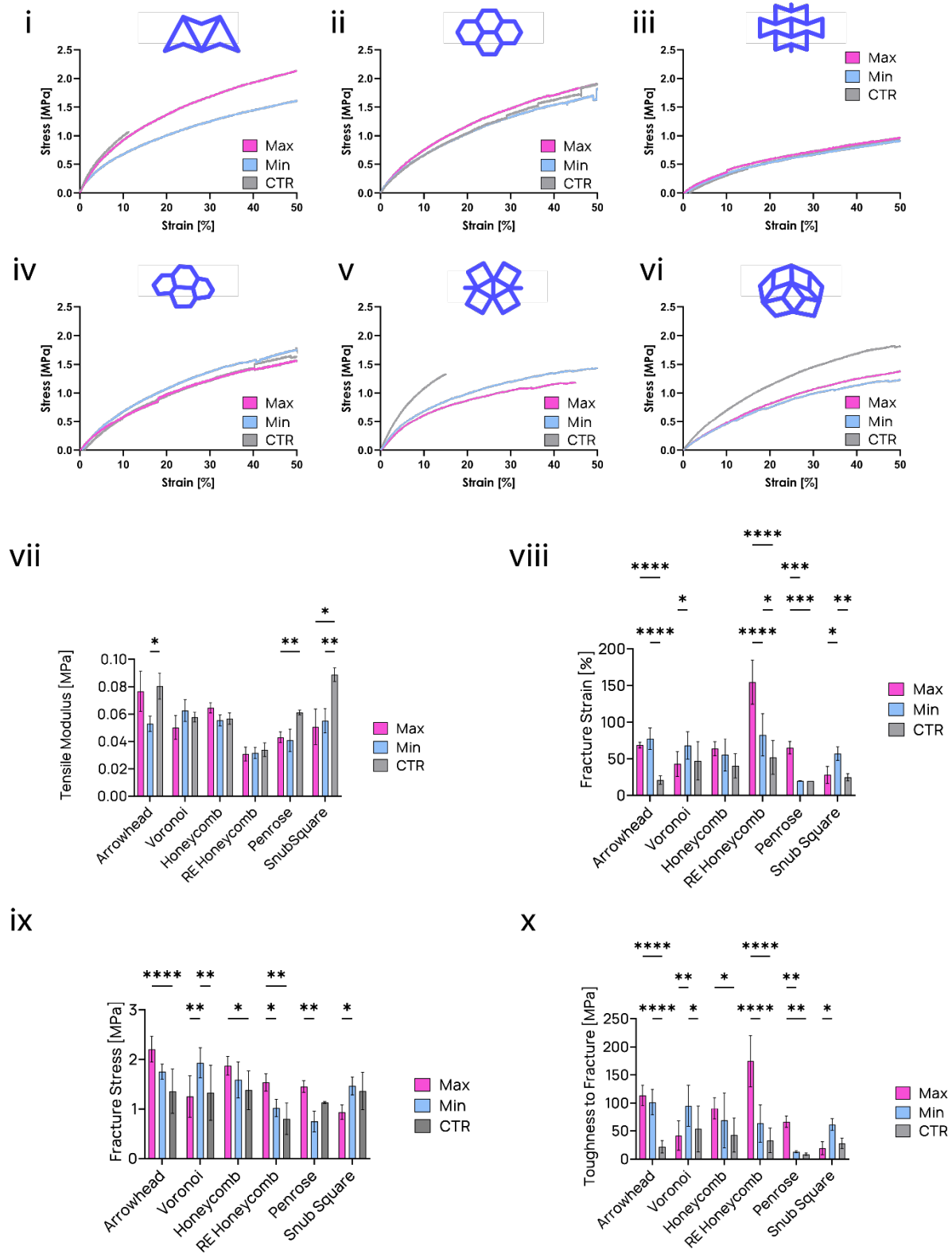

**Figure 5. Mechanical response to uniaxial tensile loading of the six geometries fabricated using the conventional or optimized printing path design. (i-vi)** Stress-strain curves for the arrowhead (i), honeycomb (ii), re-entrant honeycomb (iii), Voronoi (iv), snub-square (v) and Penrose tiling (vi). Curves represent average values from five replicates. **(vii-x)** Tensile modulus (vii), fracture strain (viii) and stress (ix) and toughness to fracture (x) values for all geometries and fabrication methods, plotted as mean ± standard deviation. Figure abbreviations: MAX: Maximum-based Optimization, MIN: Minimum-based Optimization , CTR: Control. Statistical significance: **\***: P<0.05, **\*\***: P<0.01, **\*\*\***: P<0.001**, \*\*\*\*:** P<0.0001**.**

## Shape Formability Under Out-Of-Plane Loading

The effect of the printing path optimization on mechanical performance under out-of-plane loading was assessed using ball-burst tests (**Figure 6**). The load-displacement plots revealed a stiffer response of the control conditions across all geometries with the exception of the honeycomb design (**Figure 6 (i-vi)**). Both minimum and maximum optimized conditions showed a similar response to out-of-plane loading and appeared to almost coincide up to at least 18 mm of displacement in all designs.

Additionally, the bursting strength of each condition was assessed, to investigate how the printing trajectory affects the maximum load that the lattices can withstand before the onset of mechanical failure (**Figure 6 (vii)**). Compared to the control samples, the maximum-based optimization increased the bursting strength of the Voronoi and honeycomb conditions 1.5- and 1.2-fold, respectively, while no significant differences were observed for the rest of the conditions. Additionally, differences between the optimized constructs were observed for the snub square lattices, with the maximum-based optimization increasing the bursting strength to 126.47±7.13 N compared to 77.65±4.12 N, which was the bursting strength of the minimum-based optimization. Conversely, the minimum based optimization yielded a slightly higher bursting strength compared to the maximum-based optimized condition (103.59±7.95 N and 91.066.97 N, respectively), which, however, was not significantly different from the control (108.93±4.61 N).

Despite the differences in ultimate strength, bursting displacement values did not show considerable variations among the optimized and non-optimized conditions across most geometries (**Figure 6 (viii)**). Notably, significant reduction in the bursting displacement was measured for the minimum optimized snub square lattices (14.49±0.34 mm), compared to the maximum optimized condition (22.38±1.22 mm). Overall, it was found that differences in the printing path generation affect the formability of architected lattices and their mechanical responses during out-of-plane loading.

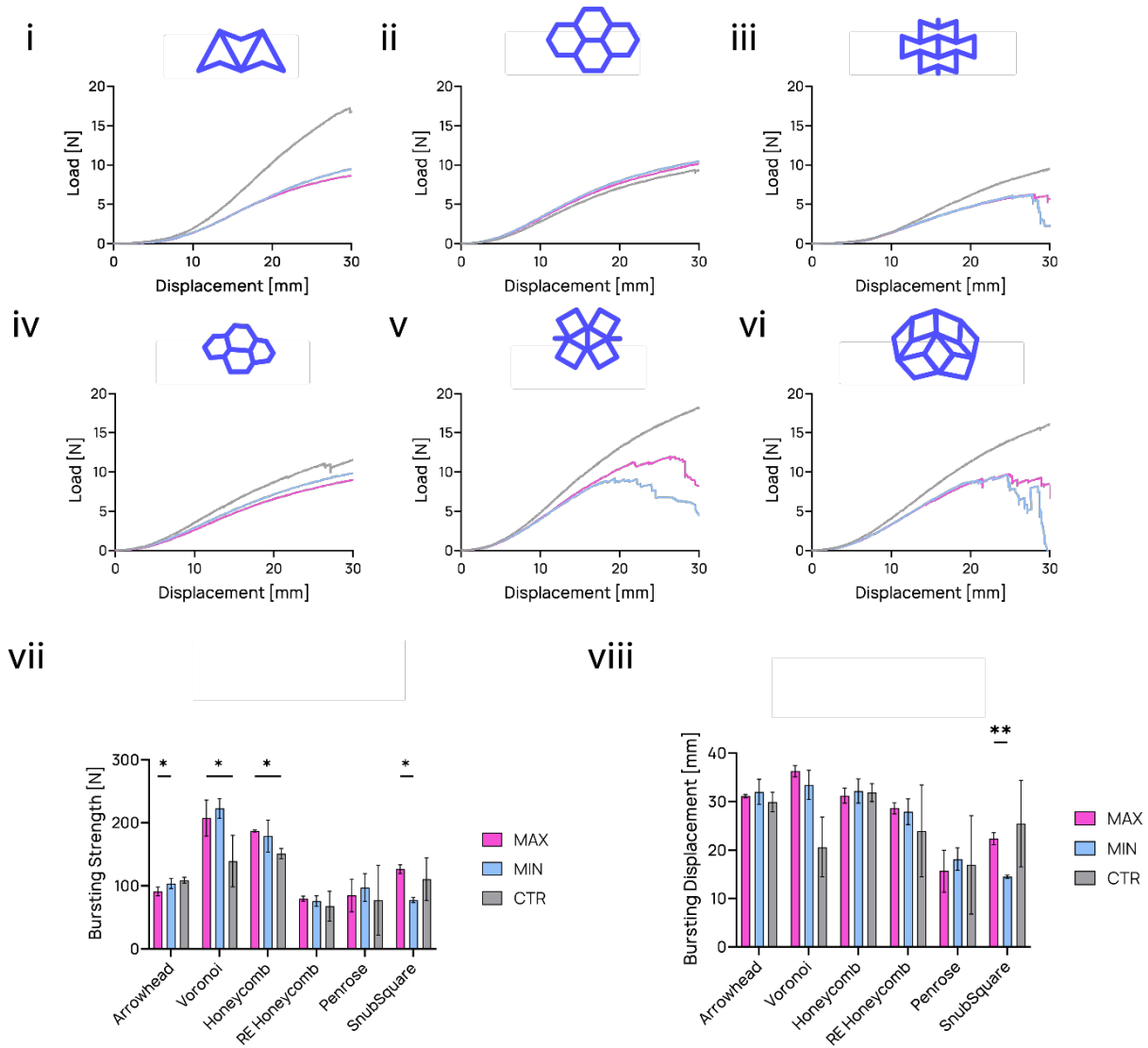

**Figure 6. Mechanical response to out-of-plane deformation of the six geometries fabricated using the conventional or optimized printing path design. (i-vi)** Load-displacement curves for the arrowhead (i), honeycomb (ii), re-entrant honeycomb (iii), Voronoi (iv), snub-square (v) and Penrose tiling (vi). Curves represent average values from five replicates. **(vii-viii)** Bursting strength (vii) and displacement (viii), plotted as mean ± standard deviation. Figure abbreviations: Max: Maximum-based Optimization, Min: Minimum-based Optimization, CTR: Control. Statistical significance: **\***: $P<0.05$, **\*\***: $P<0.01$.

## Additional features

The 3D-pneumatically-extruded lattices are presented in **Figure 7 (i)**. Our platform succeeded in generating G-codes compatible with the specific 3D printing apparatus, starting from the same optimized trajectories presented previously. The single-line trajectories were precisely followed by the CNC-based machine, resulting in the exact same movements of the FDM 3D printer. Some minimum-optimized structures, like the Voronoi and the Penrose, presented a higher oozing incidence compared to their maximum-based counterpart, as notable by the extra strings within the lattice. This effect can be easily correlated to the LTS analysis previously reported, in which it was evidenced that the maximum-based approach generally results in a higher coverage of the grid by long segments, hence a lower number of starts and stops during fabrication, and thus probability of oozing.

The concave cylinder is presented in **Figure 7 (ii-iii)** and proves how GIPPO can handle several different layers to be optimized independently, thus resulting in a different trajectory for each layer. Finally, **Figure 7 (iv)** shows a photo-realistic render of the non-planar features of our slicing platform. The

Honeycomb lattice was projected onto a spherical surface, thus proving the possibility of generating layered structures which can adapt to non-planar and non-linear printing surfaces.

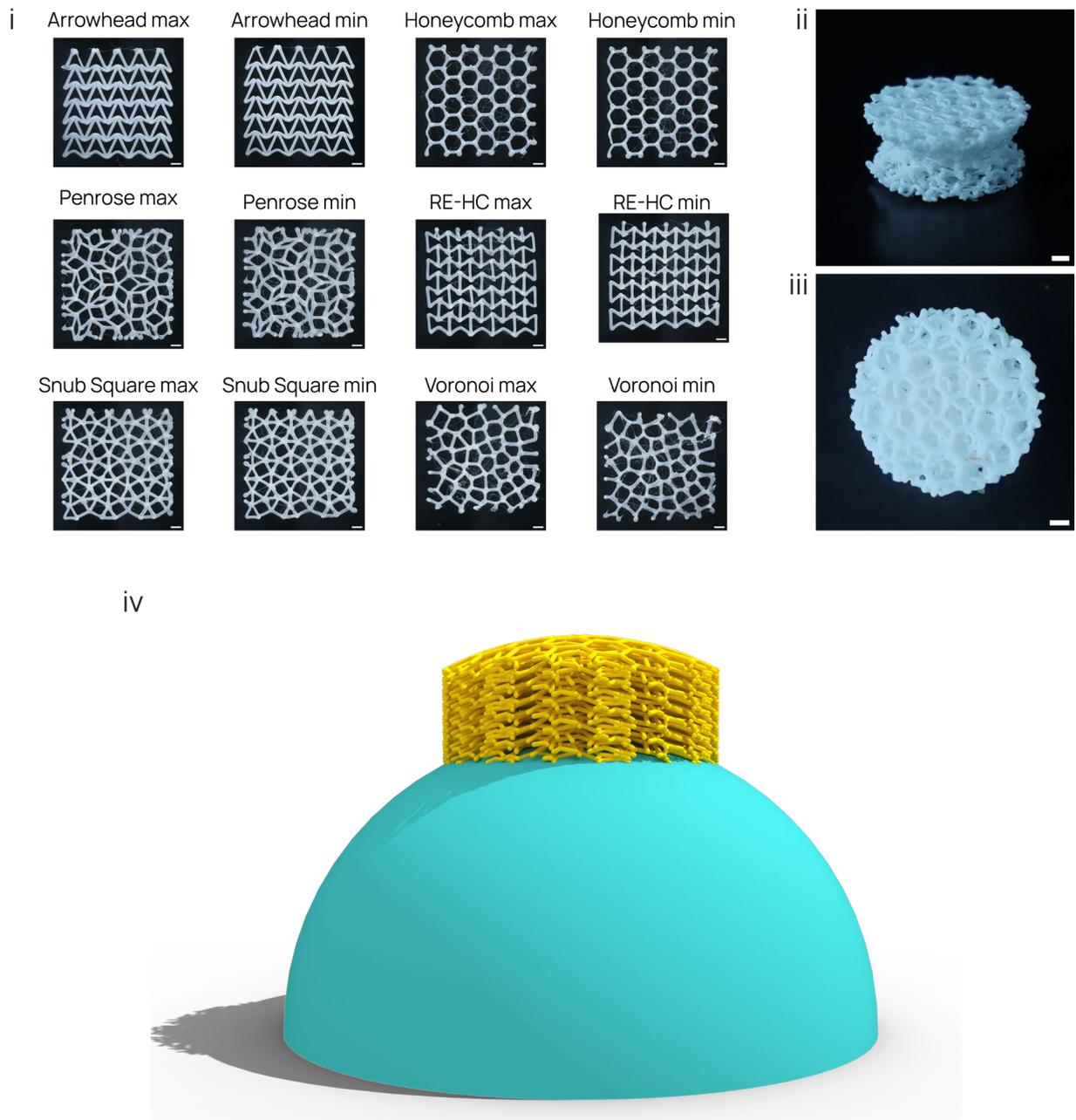

**Figure 7. Additional features of the GIPPO platform. (i)** Optimized lattices processed and fabricated through a pneumatic-based printer, showing the universality of the G-code generator. Scale bars: 1 mm **(ii-iii)** Honeycomb-filled concave cylinder. Scale bars: 1 mm. **(iv)** 3D rendering of a scaffold projected and built onto a spherical surface, highlighting the non-planar capabilities of GIPPO.

## Discussion

In this work we have presented GIPPO, an enabling platform for the printing trajectory optimization of architected lattices. We have validated the superior performance of the GIPPO-optimized constructs compared to conventionally printed lattices, using local thickness mapping, uniaxial tensile tests and out-of-plane ball-burst tests. Our platform provides a foundational and flexible framework, which can be further enriched due to its open-source nature. This collaborative model allows users to choose from already existing lattice designs as well as contribute to the expansion of the atlas with new geometries.

Six geometries were selected for their many applications in a host of fields, such as aerospace [21, 22], construction/architectural engineering [23, 24] and regenerative medicine field [2, 25, 26], as well as for the possibility to tailor their mechanical response to uniaxial and out-of-plane loadings [16]. Furthermore, these topologies can sometimes be found in different proprietary software packages, though an open platform comprising them all and allowing for a high degree of customization is still missing [16] [5]. GIPPO enhances traditional slicing software performances through the implementation of graph theory, aiming to minimize structural defects deriving from a lack of precision during the printing process. This is crucial when aiming to advanced biomimicry and specific behaviours (e.g., auxeticity, anisotropy) for which short paths and weak points can be critical. Since almost any infill geometry, either traditional or more complex one, can be represented as a graph, our best-path problem can be solved through graph theory, by taking advantage of the multi-directional path growing approach of Prim's algorithm. Our software might benefit of additional features in the future, such as post-optimization merging, which could result in a greater minimization of small paths. In particular, paths with a common extreme can be merged into one. Furthermore, future updates will include spline-based geometries as an addition to the polygonal ones currently in the atlas. Since every segment is associated with its length as an equivalent of the graph edge weight, these topologies could be implemented by upgrading the conversion process after the optimization. As of now, such lattices can be achieved by increasing the number of segmentation points for each segment.

Our algorithm demonstrated robustness on results quality, although being random-based. At the same time, the optimization process itself was proven to be effective, showing the qualitative (i.e., path reconstruction) and quantitative (i.e., OE analysis) differences between high- and low-score results. Additionally, the optimized trajectories resulted in shorter printing times, with reductions ranging from 6% to 35% (depending on the geometry and the optimization method) compared to the control trajectories, as presented in **Table S3**. Structures with a higher number of short segments (i.e., a smaller LTS), such as the Voronoi, made an exception with slower printing processes. This can be explained by one of the features of our software, which adds slow movements in case of short segments (i.e., 60% of the original speed), as shown in **Video S1**. This allows for a more precise extrusion of short paths, and hence better adhesion points between segments. Depending on many factors, such as the pores geometry, and either the tuning or removal of the slower short paths feature, fabrication times of GIPPO-optimized lattices can be enhanced even more. At the same time, as shown in the total path length analysis, the optimization brought by our software can save printing material, and decrease the cost of fabrication [27] [28] [29].

Substantial differences between the printing fidelity of the optimized and control conditions were observed, with the conventionally fabricated lattices demonstrating local shape deviations, excess material deposition and missing struts. The reconstruction of the printed paths validated this observation by revealing a 2-fold increase in the total length travelled by the printing head during the lattice fabrication for the arrowhead, re-entrant honeycombs, Snub Square and Penrose geometries. Experimentally, this translates to more material deposition and, hence, thicker struts. While the honeycomb and Voronoi path lengths were not significantly affected by the printing path generation strategy, notable deviations from the original design were observed in these conditions as well. This can be explained by the tendency of commercial 3D printing programs to prioritize external surface quality towards bulk mechanical properties. For this reason, printed structures are typically made with external walls of the highest quality offered by the system, resulting in a minimization of motion of travel and, thus, crossing of already printed parts, oozing, and stringing. As a consequence, eventual over- and/or under-extrusion phenomena are camouflaged in the G-code as motions. Given the typical applications of commercial slicing software (*i.e.,* large-scale functional objects, hobbies, household tools, etc.), such features maximize the printing outcome. However, when used to print lattice-like structures, these optimization strategies can be counterproductive.

The structures we fabricated are "single filament"-based, and in most case slicing engines are not optimized for micro-scale geometrical accuracy and filament-to-filament adhesion. This explains the re-processing of the lattices evidenced in the reverse-engineered trajectories analysis. Since our geometries are not built-in infills for commercial software, the only way to process them in conventional ways is to force the slicer to see them as contours to be filled (if there is enough empty space between contour walls) with available infills (*e.g.,* woodpile, grids, etc.). Consequently, the 3D models were adjusted to have adjacent segments tangent in one point, and thus higher incidence of inaccuracy and empty spots in the lattice. This is a major downside towards the fabrication of advanced lattices, where precision is crucial and imperfections could propagate over different layers resulting in a loss of functionality. Furthermore, while GIPPO allows for the design of samples for tensile testing (i.e., by adding handles at the extremes) *in house*, this does not happen with traditional slicers. Handles have to be generated and placed manually, thus depending on the user experience and increasing the risk of failure at the lattice-handle interface. Our platform guarantees a smooth integration between the two components, thus the risk of unexpected mechanical failures.

Another main difference between GIPPO-optimized and control structures can be found in the final printing quality. Optimized results show a more pronounced oozing effect compared to the control ones. This depends on the way printing paths are sorted. In the case of GIPPO, after the optimization, the paths are optimized from longest to shortest, based on the number of segments of the path. This might result in consecutive paths being distant from one another, and thus in long non-printing movements. On the other hand, towards the aforementioned goal of increasing bulk quality and properties, commercial slicers tend to sort printing paths on a distance-basis, minimizing the non-printing distance. In this context, GIPPO structures could be enhanced by introducing user-defined optimization and/or sorting directions, which could also be tailored according to the specific application. This consistent thickness deviation observed when comparing the CAD designs to the GIPPO-optimized lattices could also be attributed to the die swell effect, which is inherent to extrusion-based additive manufacturing. This phenomenon occurs when a polymer melt is extruded through a nozzle and the diameter of the printed fibre is greater than the inner diameter of the nozzle, as a result of the elastic recovery of the material [30]. Specifically, in our study, the lattices were fabricated using thermoplastic polyurethane. Elastic materials are known to exhibit a more pronounced die swell effect due to their viscoelastic nature when melted. Therefore, the use of stiffer polymers could result in lattices with average strut thickness values closer to the theoretical ones, as they will be more resistant to the elastic deformation that causes the expansion. As a future feature of the GIPPO platform, the expected swell ratio of the extruded material could be estimated based on the printing parameters and the rheological characteristics of the extruded material [31]. This would result in theoretical thickness values that better illustrate the experimental conditions. In such case, the optimized lattices are expected to demonstrate an even lower deviation from the expected local thickness values.

Additionally, the possibilities offered by the GIPPO platform could be further complemented by different extrusion printing modalities, including multimaterial gradients, embedded printing, chaotic printing, as well as by increasing the degrees of freedom (DOF) of the fabricating system (e.g., robotic printing). For instance, some attempts have been done towards multi-DOF systems for *in vivo* biofabrication applications [32, 33]. However, the lack of a trajectory generation platform to accommodate the non-planar and volumetric capabilities of such extrusion system limits the great potential of them. Given the versatility of GIPPO to optimize any geometry once divided into nodes and edges, the process could be easily extended to 3D lattices and structures. In addition to that, GIPPO's non-planar capabilities to generate complex architectures onto any regular or irregular surface could enhance *in vivo* biofabrication applications (e.g., adapting a 3D scaffold to physiological surfaces). Furthermore, the complexity of the architected lattices could be additionally enhanced by implementing advanced extrusion modes, *e.g.*, harnessing of fibre coiling and instability [34]. Moreover, our optimization approach could have significant benefits for bioprinting applications. By optimizing the printing trajectories, biofabrication could be completed faster, reducing the total time cell-laden bioinks need to stay in the cartridge. This could improve the overall viability of the encapsulated cells. Looking beyond extrusion, the implementation of our platform could be extended to other additive manufacturing methods that rely on trajectory-based fabrication approaches, such as selective laser sintering and laser metal deposition. All these features might be easily implemented in the future given the firmware-agnostic nature of the GIPPO platform, meaning that it can generate codes compatible with different machines from the same optimized design.

In this study we showcased the effect of the GIPPO optimization on six architected lattices of significant structural complexity. Despite the observation of several clear trends, selecting an optimization approach as the optimal one would require a considerably larger amount of data. For this purpose, the atlas of architected designs could be expanded and subsequently optimized by the GIPPO platform. This could be feasible by allowing users to share the printing optimization results of their own lattice designs on an open library. It should, additionally, be noted that as long as the printing fidelity of the lattice is satisfactory for the intended application, the optimization strategy could also depend greatly on the application itself. For instance, the printing fidelity of the snub square lattices was similar for both minimum and maximum-based optimization methods. However, a higher fracture strain, fracture stress and toughness were exhibited by the minimum-based optimized conditions, under uniaxial tensile loading. On the contrary, the maximum-based optimized samples showed a higher bursting strength and displacement during out-of-plane tests. Therefore, the user could choose the printing trajectory that better suits the end-application, depending on the expected loading conditions.

In order to enhance even more the design generation process of GIPPO and the overall user experience, the aforementioned choice could be automated in the future and derived from literature analysis. In particular, given the desired behaviour and application of interest, the software could suggest the best unit cell, pores dimension, strut thickness and optimization method. Such values might be supported by experimental and simulated data, uploaded by both the developers and the users, thus creating an open community of researchers contributing to the growing field of fabrication of complex architectures.

## Conclusion

GIPPO is a novel, open-source slicing platform for the printing path optimization of architected lattices that can be used with different printing systems, materials, designs and accommodate both planar and non-planar printing, addressing a gap in commercial slicers. We have demonstrated that software-inherent lattice fabrication hurdles, including local thickening, multiple short segments and oozing, can be substantially minimized by GIPPO, which yields constructs of high fidelity. Moreover, we identified a direct effect of the printing path generation on the mechanical performance of the fabricated lattices during both uniaxial tension and out-of-plane loading. Our work opens new opportunities for the fabrication of complex, architected lattices for use in a host of fields including tissue engineering, biofabrication, aerospace, construction and automotive research.

## Funding

This publication is part of the project 3D-MENTOR (with project number 18647) of the VICI research programme, which is financed by the Dutch Research Council (NWO).

## Conflict of Interest

The authors of this manuscript claim that they have no competing interests.